\documentclass[aps,prx,twocolumn,showpacs,superscriptaddress,amsmath,amssymb,amsfonts,floatfix,longbibliography]{revtex4-1}

\usepackage{graphicx}
\usepackage{placeins}
\usepackage{float}
\usepackage{dcolumn}
\usepackage{bm}
\usepackage{amssymb}
\usepackage[version=4]{mhchem}
\usepackage{microtype}
\newcolumntype{C}[1]{>{\centering\arraybackslash}p{#1}}
\usepackage{xfrac}
\usepackage{gensymb}
\usepackage{xcolor}
\usepackage{multirow}
\usepackage{comment}
\usepackage{times}
\usepackage[varg]{txfonts}
\usepackage{mathrsfs}
\usepackage{amsmath,amssymb,bm,mathtools}
\usepackage{amsfonts}
\usepackage{booktabs}
\usepackage{natbib,hyperref}
\hypersetup{
	colorlinks=true, 
	linkcolor=blue,  
}
\usepackage{soul}
\usepackage{epstopdf}
\usepackage{array}
\usepackage{multirow}
\usepackage{tcolorbox}
\usepackage[percent]{overpic}
\usepackage{bbm}

\usepackage{widetable}
\usepackage{array}
\usepackage{xcolor,colortbl}
\usepackage{lipsum}

\begin{document}

\title{Dynamic Parallel Spin Stripes from the 1/8 anomaly \\ to the End of Superconductivity in La$_{1.6-x}$Nd$_{0.4}$Sr$_x$CuO$_4$}

\author{Qianli Ma}
\affiliation{Department of Physics and Astronomy, McMaster University, Hamilton, Ontario, L8S 4M1, Canada}

\author{Evan M. Smith}
\affiliation{Department of Physics and Astronomy, McMaster University, Hamilton, Ontario, L8S 4M1, Canada}

\author{Zachary~W. Cronkwright}
\affiliation{Department of Physics and Astronomy, McMaster University, Hamilton, Ontario, L8S 4M1, Canada}

\author{Mirela Dragomir}
\affiliation{Brockhouse Institute for Materials Research, Hamilton, Ontario, L8S 4M1, Canada}
\affiliation{Electronic Ceramics Department, Jožef Stefan Institute, 1000 Ljubljana, Slovenia}

\author{Gabrielle Mitchell}
\affiliation{Department of Physics and Astronomy, McMaster University, Hamilton, Ontario, L8S 4M1, Canada}

\author{Alexander~I.~Kolesnikov}
\affiliation{Neutron Scattering Division, Oak Ridge National Laboratory, Oak Ridge, TN 37830, United States}

\author{Matthew B. Stone}
\affiliation{Neutron Scattering Division, Oak Ridge National Laboratory, Oak Ridge, TN 37830, United States}

\author{Bruce~D.~Gaulin}
\affiliation{Department of Physics and Astronomy, McMaster University, Hamilton, Ontario, L8S 4M1, Canada}
\affiliation{Brockhouse Institute for Materials Research, Hamilton, Ontario, L8S 4M1, Canada}
\affiliation{Canadian Institute for Advanced Research, MaRS Centre, West Tower 661 University Ave., Suite 505, Toronto, ON, M5G 1M1, Canada}

\date{\today}


\begin{abstract}
We have carried out new neutron spectroscopic measurements on single crystals of La$_{1.6-x}$Nd$_{0.4}$Sr$_x$CuO$_4$ from 0.12 $\le x \le$0.26 using time-of-flight techniques.  These measurements allow us to follow the evolution of parallel spin stripe fluctuations with energies less than $\sim$ 33 meV, from $x$=0.12 to 0.26.  Samples at these hole-doping levels are known to display static (on the neutron scattering time scale) parallel spin stripes at low temperature, with onset temperatures and intensities which decrease rapidly with increasing $x$.  Nonetheless, we report remarkably similar dynamic spectral weight for the corresponding dynamic parallel spin stripes, between 5 meV to 33 meV, from the 1/8 anomaly near $x$=0.12, to optimal doping near $x$=0.19 to the quantum critical point for the pseudogap phase near $x$=0.24, and finally to the approximate end of superconductivity near $x$=0.26.  This observed dynamic magnetic spectral weight is structured in energy with a peak near 17 meV at all dopings studied. Earlier neutron and resonant x-ray scattering measurements on related cuprate superconductors have reported both a disappearance with increasing doping of magnetic fluctuations at ($\pi$, $\pi$) wavevectors characterizing parallel spin stripe structures, and persistant paramagnon scattering away from this wavevector, respectively.  Our new results on La$_{1.6-x}$Nd$_{0.4}$Sr$_x$CuO$_4$ from 0.12 $\le x \le$0.26 clearly show persistent parallel spin stripe fluctuations at and around at ($\pi$, $\pi$), and across the full range of doping studied.  These results are also compared to recent theory. Together with a rapidly declining x-dependence to the static parallel spin stripe order, the persistent parallel spin stripe fluctuations show a remarkable similarity to the expectations of a quantum spin glass, random t-J model, recently introduced to describe strong local correlations in cuprates.

\end{abstract}

\maketitle

\section{Introduction}

High temperature superconductivity in hole-doped cuprates arises
from the presence of mobile holes in the quasi-two dimensional CuO$_2$ layers, introduced by chemical doping.  The un-doped parent material for the 214 family of superconductors is La$_2$CuO$_4$, a Mott insulator with three dimensional antiferromagnetic order at $\sim$ 300 K\cite{Kastner,keimer1992magnetic}.  On doping La$^{3+}$ with Sr$^{2+}$ in either La$_{2-x}$Sr$_x$CuO$_4$ (LSCO) or La$_{1.6-x}$Nd$_{0.4}$Sr$_x$CuO$_4$ (Nd-LSCO), or with Ba$^{2+}$ in La$_{2-x}$Ba$_x$CuO$_4$ (LBCO), three dimensional antiferromagnetism is rapidly destroyed, and replaced, by $x$ $\sim$ 0.02, with incommensurate, quasi-two dimensional static magnetism which onsets at much lower temperatures\cite{Kastner,keimer1992magnetic,Kastner,3dAF2,3dAF3,3dAF4}.  

Neutron scattering famously sees this quasi-two dimensional, incommensurate spin order as elastic Bragg peaks, with finite reciprocal space widths, or inverse correlation lengths, in the ab basal plane, \cite{tranquada1995evidence,tranquadaorderparameter,tranquada1999glassy,moment}, which can be referred to as quasi-Bragg peaks.  Using a tetragonal unit cell, for which $a$=$b$=3.88 $\AA$, diffraction from diagonal spin stripe order is first observed at low doping with quasi-Bragg peaks at four positions split off from the ($\frac{1}{2}$, $\frac{1}{2}$, 0), or ($\pi$, $\pi$, 0), position in reciprocal space\cite{tranquada1995evidence}.  Wakimoto and collaborators made the remarkable discovery that the incommensurate pattern of quasi-Bragg peaks in LSCO rotates within the $a^*$-$b^*$ plane of reciprocal space by 45 $\degree$, to ordering wavevectors of the form ($\frac{1}{2}$, $\frac{1}{2}\pm\delta$, 0) and ($\frac{1}{2}\pm\delta$, $\frac{1}{2}$, 0), forming parallel spin stripe order, at $x$=0.05, which is also the onset of superconducting ground states in LSCO\cite{diagonalstripe}.  A similar rotation between so-called diagonal and parallel spin stripe magnetism has also been observed in LBCO and Nd-LSCO\cite{dunsigerlbco,Parallelstripe,wakimotondlsco,Kyle}.

\begin{figure}[tbp]
\hspace*{-0.2in} \includegraphics[width=3.5in]{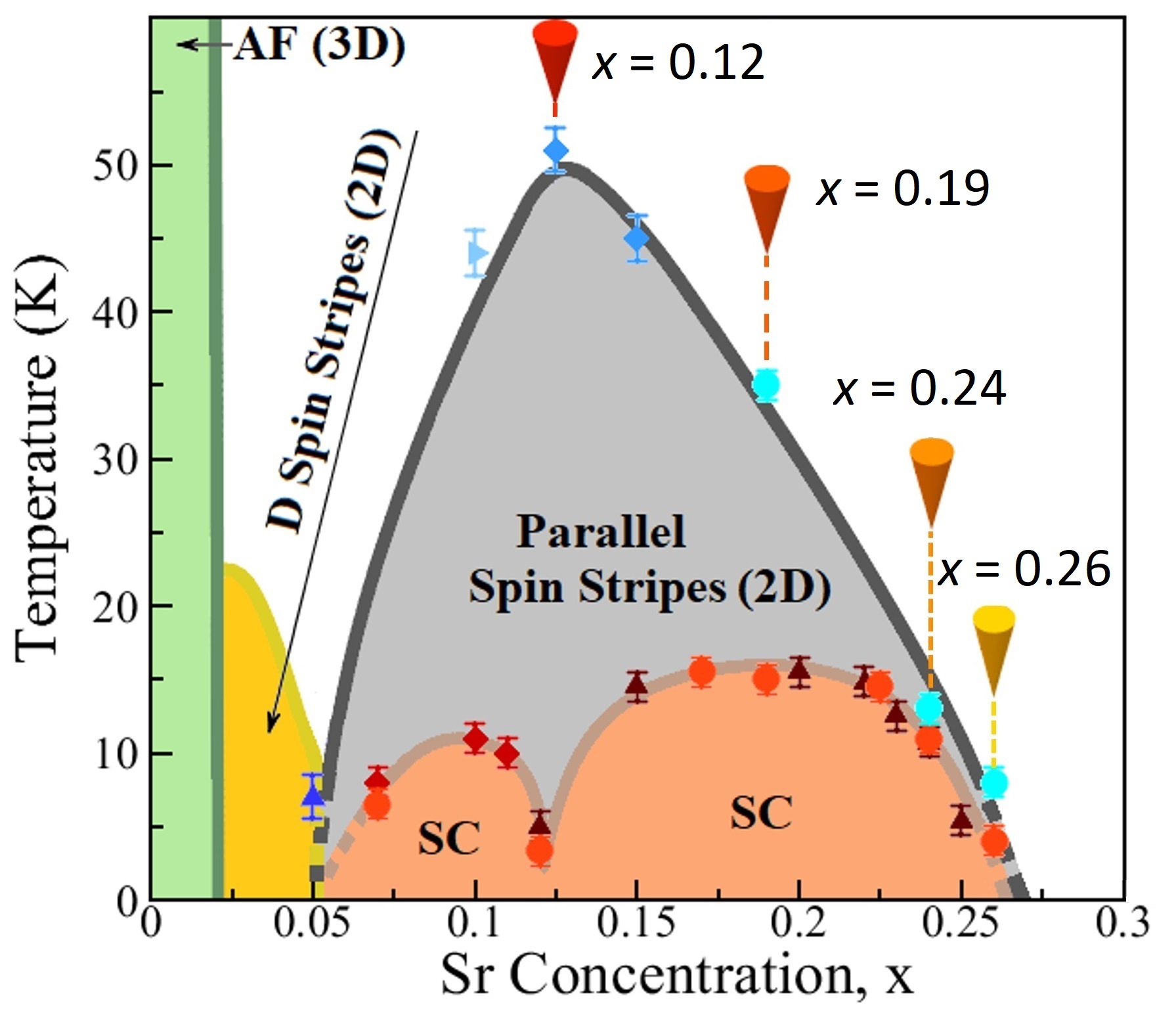}
\caption{The magnetic and superconducting phase diagram for Nd-LSCO is shown, with markers indicating the hole-doping level $x$ = 0.12, 0.19 0.24 and 0.26 of the single crystal samples which are the subject of the present inelastic neutron scattering study. This plot is modified from Ref \cite{Kyle}  , where the references for the magnetic and superconducting transition temperatures are given.}
\label{phasediagram}
\end{figure}

Recently, much interest has focused on Nd-LSCO as it possesses the full complexity of the hole-doped cuprate phase diagram while exhibiting relatively low superconducting transition temperatures.  The low temperature magnetic and superconducting phase diagram for Nd-LSCO is summarized in Fig. \ref{phasediagram}.  The low superconducting $T_C$'s allow for relatively easy access to normal state properties at low temperatures, as superconductivity can be destroyed with application of relatively modest, and therefore practical, magnetic field strengths.  This in turn has allowed detailed studies of both electrical and thermal transport in normal state Nd-LSCO at all doping levels.  In particular it allowed recent thermodynamic measurements which reveal a quantum critical point at $p^*$=0.23, signifying the end of the elusive pseudogap phase\cite{michon2019thermodynamic}. Hall effect measurements show that $p^*$ is coincident with an abrupt change in the Hall number, n$_H$, from being proportional to $p$, to be proportional to 1+$p$, consistent with a dramatic change in Fermi surface coincident with $p^*$\cite{badoux2016change,collignonprb,lizaire2020transport}.  The Nd-LSCO materials system can be grown as high quality, large single crystals, appropriate for neutron scattering measurements, over a relatively large range of hole-doping, $x\le$0.26, in La$_{1.6-x}$Nd$_{0.4}$Sr$_x$CuO$_4$\cite{mirelakyle}. 

Recent elastic neutron scattering measurements have shown static, quasi-two dimensional parallel stripe order in Nd-LSCO single crystals with $x$=0.12, 0.19, 0.24 and 0.26\cite{Kyle}. This order is not true long range three dimensional order, but corresponds to large in-plane correlation lengths in the $a$-$b$ plane, and short correlation lengths along $c$, such that well-defined incommensurate quasi-Bragg peaks are easily observed at low temperatures.  With finite correlation lengths at all temperatures and spin 1/2 degrees of freedom, such a system also corresponds to a quantum spin glass.  The static (on the time scale of the neutron) magnetic moment associated with the parallel stripe order is a small fraction of the 1 $\mu_B$ expected for $S$=1/2 Cu$^{2+}$ ions, and the intensity of the quasi-Bragg peaks fall off with increased doping, while the in-plane correlation lengths correspondingly decrease. Nonetheless, they remain observable even at $x$=0.26, which is close to the end of the superconducting dome, and which displays a high temperature tetragonal phase down to at least $T$=2 K.  This work is broadly consistent with recent NMR studies on the LSCO where magnetism has been observed at higher doping than previously believed \cite{frachet2020hidden}. 

Resonant elastic x-ray scattering performed on samples cut from the same Nd-LSCO single crystals studied here also show clear evidence for the corresponding static parallel {\it charge} stripe Bragg peaks, at $x$=0.12 and 0.17, but not at $x$=0.19\cite{david2020vanishing} above T=22 K.  
In LSCO, recent scattering have reported both an end to parallel charge stripes, or charge density wave (CDW) scatterings at $x$ $\sim$ 0.18, and the observation of CDWs \cite{cdw2miao2021charge} to $x$ = 0.21. 

Modern time-of-flight (TOF) neutron scattering techniques allow comprehensive four dimensional (4D) measurements, covering three reciprocal space dimensions as well as energy.  However, measurements with full 4D sensitivity have only been employed relatively recently. In the past, TOF neutron scattering performed on direct-geometry chopper instruments were often performed with single crystals of cuprate superconductors in a single orientation, with $k_i$ // $c$-axis.  Information on the evolution of the dynamic magnetic spectral weight within a tetragonal basal plane of reciprocal space, $a^*$-$b^*$, could be obtained, but under these conditions, the energy transfer of the scattering is conflated with the $c^*$ dependence of the scattering.  This could be used to good effect provided that there either was no $c*$-dependence to the scattering, or that the $c*$-dependence of this scattering was well understood.  However, the hole-doped cuprates are complex oxides with many atoms per unit cell, and hence with complex phonon spectra, as well as crystalline electric field excitations in the case of Nd-LSCO, in addition to magnetic excitations. Hence a full 4D data set and analysis is preferred.  

In this paper, we present comprehensive 4-D TOF neutron scattering data on single crystal Nd-LSCO samples of $x$ = 0.12, 0.19, 0.24 and 0.26, covering the under-doped, optimally-doped and over-doped regions of the phase diagram, as indicated in Fig. \ref{phasediagram}. Our results clearly show that the dynamic parallel spin stripes persist beyond the quantum critical point at $p^*$, up to the very end of the superconducting dome with remarkably little decrease in spectral weight until an $\sim$ 33$\%$ drop in intensity is observed at $x$=0.26. We compare our results to preexisting studies of dynamic magnetic spectral weight from both neutron and resonant inelastic x-ray scattering in related 214 cuprates \cite{dean2012,dean2013persistence,Robarts2019}, as well as to recent theory based on the Hubbard model at optimal and relatively high hole-doping\cite{hubbard}, as well as to work on a random t-J model, appropriate to a quantum spin glass\cite{schackleton2021}.  

\begin{figure*}[hptb]
\hspace*{-0.3in}\linespread{1}
\includegraphics[width=7.2in]{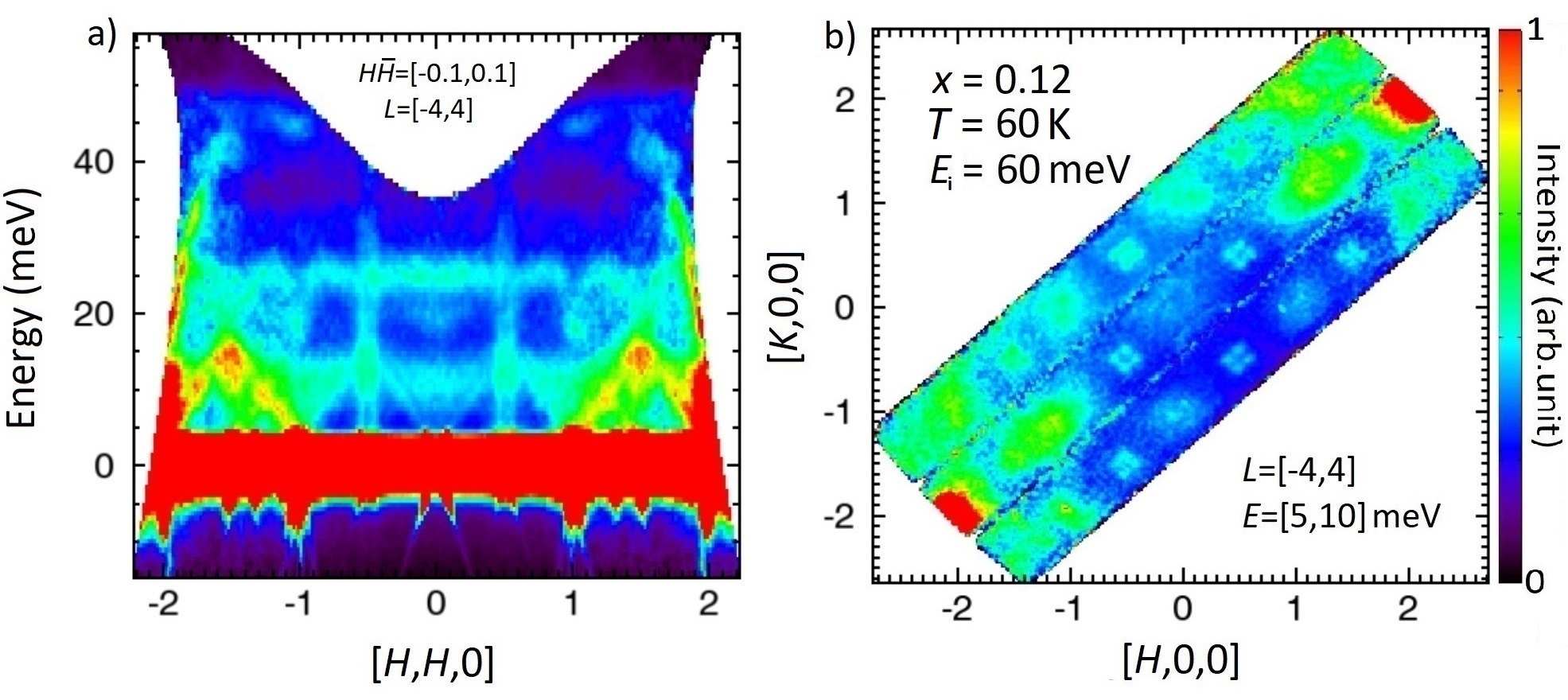}
\caption{Typical TOF neutron scattering results for the Nd-LSCO $x$ = 0.12 single crystal are shown at T=60 K. Panel a) shows an energy vs $HH$ plot, with integrations of -4 $\le$L$\le$4 and -0.1 $\le$$H,\bar{H}$$\le$ 0.1. Panel b) shows an energy slice for 5 meV $\le$ E $\le$ 10 meV, within the $HK$ plane of reciprocal space and with an $L$ integration of -4 $\le$L$\le$4. These data have had an empty sample can, background data set subtracted from it. A hybridization of optical phonon modes and the 25 meV CEF level can be observed in a). The ladder-like features centered around [+$\frac{1}{2}$, +$\frac{1}{2}$, $L$] and [-$\frac{1}{2}$, -$\frac{1}{2}$, $L$] in a) are the dynamic parallel spin stripes, which are also recognized by the quartet of peaks around [+$\frac{1}{2}$, +$\frac{1}{2}$,0], [+$\frac{3}{2}$, +$\frac{1}{2}$, 0] and equivalent wavevectors in b).}
\label{example} 
\end{figure*}

\section{Sample Preparation \& Experimental Methods}
High quality single crystals of Nd-LSCO with \textit{x} = 0.12, 0.19, 0.24 and 0.26 were grown using the traveling solvent floating zone technique at McMaster University. The resulting single crystals were of mass 3.8g, 3.6g, 4.0g, and 3.9g for each of $x$=0.12, 0.19, 0.24 and 0.26, respectively. The Nd-LSCO sample with \textit{x} = 0.19 was comprised of two co-aligned single crystals each weighs $\sim$ 1.5g and 2.1g. The single crystals were produced using a four-mirror Crystal Systems Inc. halogen lamp image furnace at approximate growth speeds of 0.68 mm/hr, and growths lasting for approximately 1 week each. The Sr concentration of each single crystal was determined by careful correlation of the structural phase transition temperatures with pre-existing phase diagrams, as described in \cite{mirelakyle}.  Further details regarding the materials preparation and single crystal growth of these samples, as well as determination of their stoichiometry, is reported in \cite{mirelakyle}.  All single crystals were scanned using a back-scattering Laue X-ray instrument to assess their single-crystal nature. Neutron diffraction measurements showed mosaic spreads of less than 0.5$\degree$ in all crystals, attesting to their high quality, single crystalline nature.

Time-of-flight (TOF) neutron chopper spectrometer measurements were carried out on all four of the \textit{x} = 0.12, 0.19, 0.24 and 0.26 single crystals using the direct-geometry chopper time-of-flight spectrometer, SEQUOIA, at the Spallation Neutron Source, Oak Ridge National Laboratory \cite{granroth2010sequoia}.  All measurements were performed using \textit{E}$_i$=60 meV neutrons, which gave an energy resolution of $\sim$ 1.2 meV at the elastic position. The single crystal samples were loaded in closed cycle refrigerators with a base temperature of 5 K, and aligned with their $HHL$ planes coincident with the horizontal plane. The single crystal samples were rotated through 360$\degree$ about the vertical axis of the instrument in 1 degree steps during the course of any one measurement, which typically required 24 hours of counting. 


\section{Experimental Results}

\subsection{Inelastic neutron scattering}

\begin{figure}[tbp]
\hspace*{-0in}\includegraphics[width=3.5in]{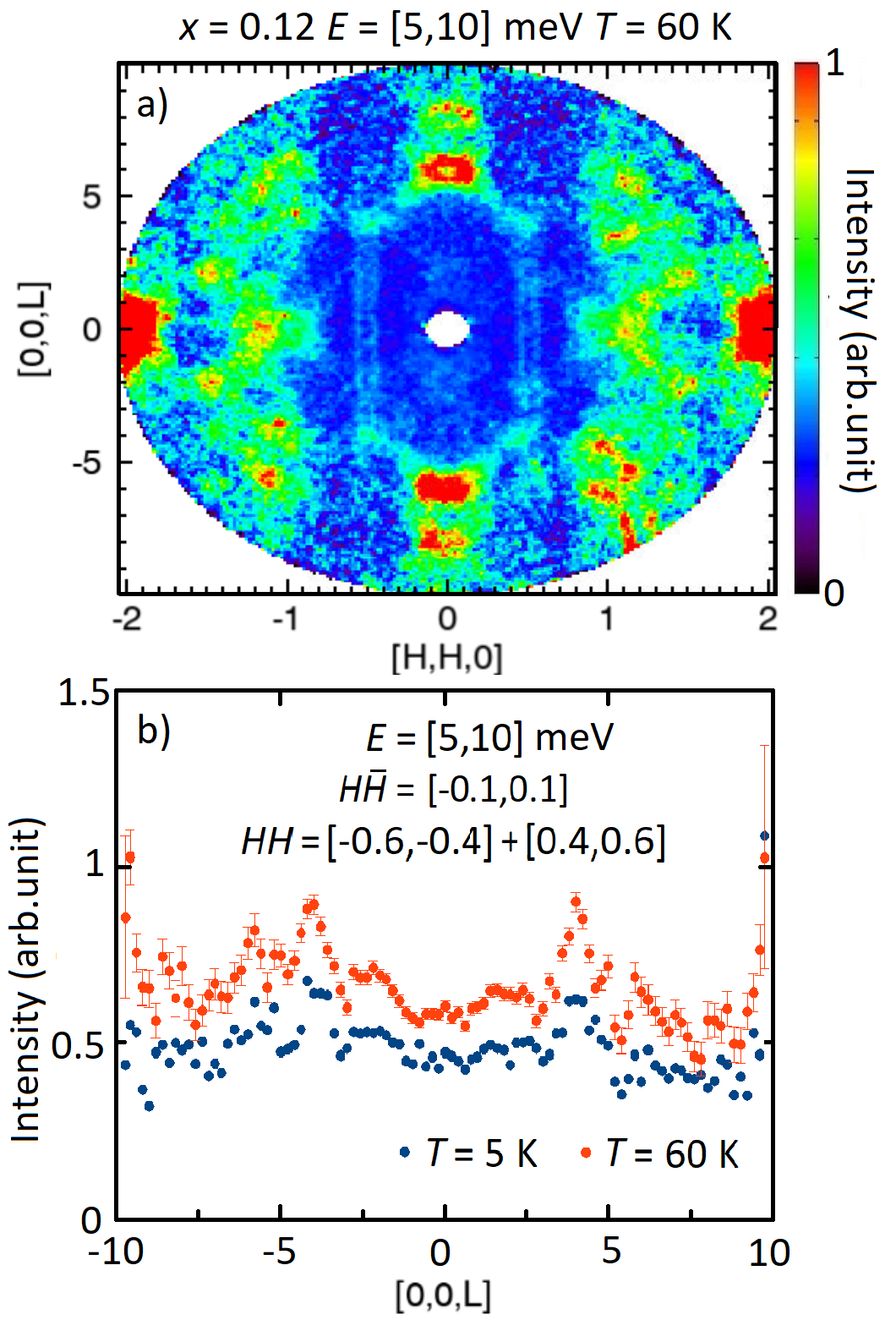}
\caption{a) A color contour map of inelastic scattering within the $HHL$ plane of the $x$ = 0.12 Nd-LSCO sample, integrated between 5 to 10 meV and $H\bar{H}$ = [-0.1, 0.1]]. This data set has had an empty sample can, background data set subtracted from it.  The dynamic parallel spin stripes can be seen as "ladder"-like features of scattering along $L$ about the $HH$ = 0.5 and -0.5 positions. b) Line scans along $L$ direction from the color contour map in a) for both 5 K and 60 K data set of $x$ = 0.12 Nd-LSCO sample. The cuts integrate data centered about both 0.5 and -0.5 in $HH$.  The weak structure at L=2, 4 and 6 is attributed to acoustic phonons, with most of the intensity due to two dimensional parallel spin stripes.}
\label{HHL} 
\end{figure}

Typical $E_i$=60 meV time-of-flight (TOF) inelastic neutron scattering data for single crystal Nd-LSCO with $x$=0.12 and T=60 K is shown in Fig. \ref{example} a) and b) and in Fig. \ref{HHL} a) and b).  This data comes from a single 4D data set, and presenting it as two-dimensional colour-contour maps, as in both Fig. \ref{example} and Fig. \ref{HHL} a), requires two integrations.  For the data shown in Fig. \ref{example} a) the plot shows $E$ (meV) vs the $H,H,0$ direction of reciprocal space, with integrations in $L$ between -4 and 4 and in $H$,$\bar{H}$,0 between $H$=-0.1 and $H$=0.1.  In Fig. \ref{example} b) the plot shows the 0,$K$,0 direction of reciprocal space vs the $H$,0,0 direction of reciprocal space, again with integrations in $L$ between -4 and 4 and in energy between 5 and 10 meV.  These data sets have had an empty sample cell background data set subtracted from them, but are otherwise plotted on full intensity scale (ie. no false zero is employed).  For the $x$=0.12 sample, measurements were taken at both $T$=5 K and 60 K.  There is little qualitative difference between these two data sets, although higher intensity is seen at energies below $\sim$ 10 meV at $T$=60 K, due to the Bose population factor.

Figure \ref{example} shows rich inelastic scattering due to the Cu$^{2+}$ spin excitations near [1/2 1/2 0] and equivalent wavevectors; both acoustic (below $\sim$ 14 meV) and optical (above $\sim$ 14 meV) phonons at relatively large $H$ and $K$; as well as Nd$^{3+}$ crystalline electric field excitations at $\sim$ 11 meV, 25 meV and 45 meV.  These are identified by the $\bf Q$ dependence of the neutron scattering cross section, which goes like the magnetic form factor squared, F(Q)$^2$, for magnetic scattering and like ($\bf \epsilon \cdot Q$)$^2$ for phonons, where $\bf \epsilon$ is the phonon eigenvector.  Nd$^{3+}$ crystal field excitations are expected to be broadly dispersionless, as they are a single ion property, and restricted to small $\bf Q$ due to the magnetic form factor. 

\begin{figure*}[tbp]
\linespread{1}
\par
\hspace*{-0.2in}\includegraphics[width=6in, height=5in]{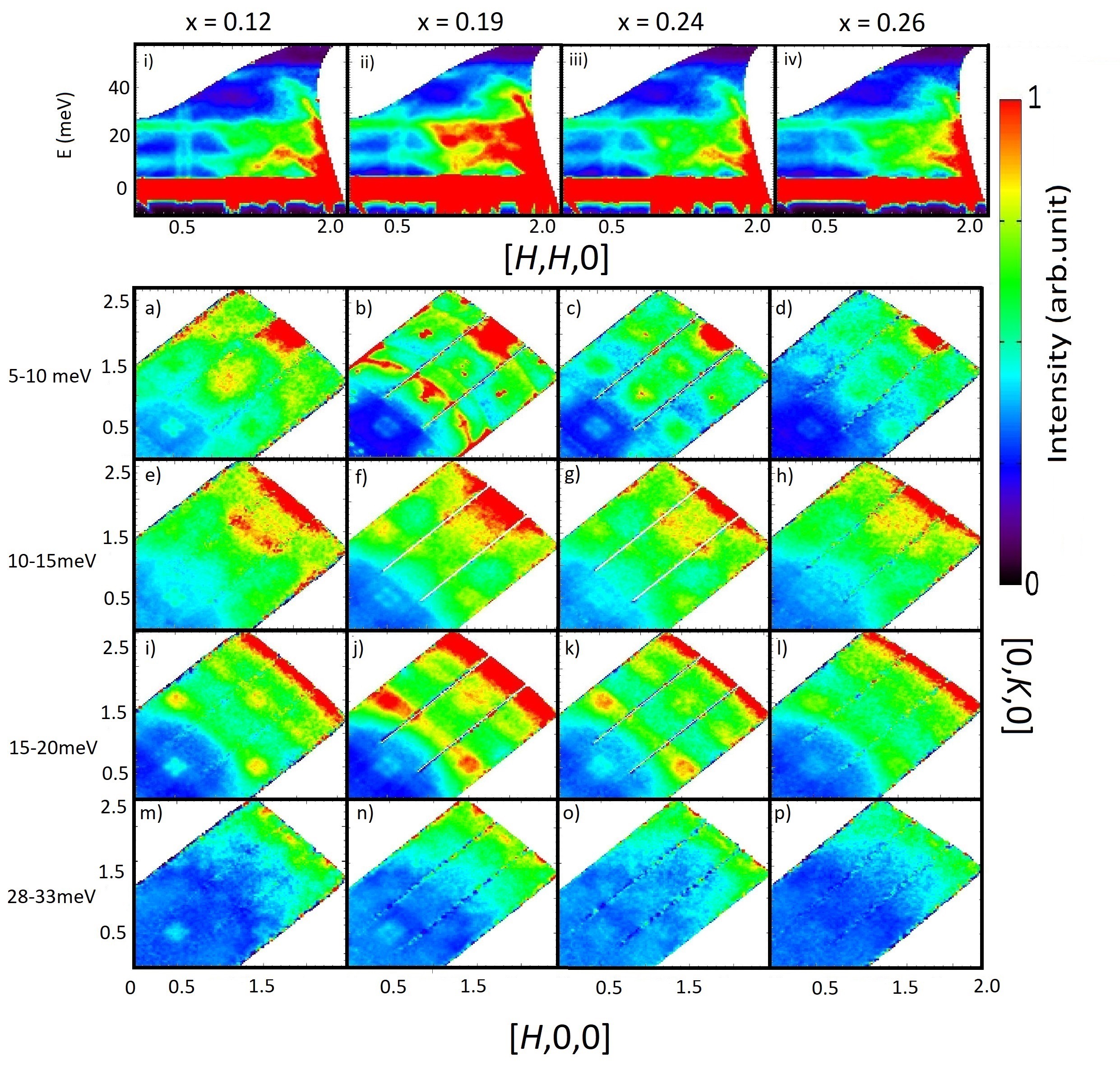}
\par
\caption{Top row: Inelastic neutron scattering shown as Energy vs $HH$ for $x$ = 0.12 i), 0.19 ii), 0.24 iii) and 0.26 iv).The rod-like feature around (0.5,  0.5, 0) represents the parallel spin stripe fluctuations. Bottom four rows, a) - p): energy cuts through the same data, from 5 to 10 meV (a to d), from 10 to 15 meV (e to h), from 15 to 20 meV (i to l), and from 28 to 33 meV (m to p) in $H,K$-maps. The incommensurate peaks around (0.5, 0.5, 0) are the same parallel spin stripe fuctuations as the rod-like features in the top row, i) to iv). The data has been normalized to a combination of Nd$^{3+}$ crystal field excitations at $\sim$ 10 and 25meV.}
\label{colorplot} 
\end{figure*}

Dynamic parallel spin stripes are clear in both Fig. \ref{example} a) and b).  In Fig. \ref{example} b) these are easily recognized as a quartet of peaks around (1/2, 1/2, 0) and equivalent ``($\pi$, $\pi$)" wavevevctors, as well as (3/2, 1/2, 0) and equivalent ``($3\pi$, $\pi$)" wavevectors.  These dynamic parallel spin stripe fluctuations are between 5 and 10 meV in energy, and are extended in $L$, hence two dimensional in nature.  Thus integrating this signal in $L$ over the low-Q range -4$\le L\le$4 is very effective.  A greater L integration allows greater intensity, but also includes more high-Q scattering originating from phonons, while a smaller L-integration provides less integrated intensity from the quasi-two dimensional magnetism. The L-dependence to the inelastic scattering between 5 and 10 meV in energy is explicitly examined below and in Fig. \ref{HHL}.  

As shown in Fig. \ref{example} a) these parallel stripe fluctuations are very dispersive, and they appear as approximate ladders around (1/2, 1/2, 0). Work on other 214 families of cuprates shows these excitations to disperse towards zone boundary energies in the 200 meV to 300 meV range\cite{tranquada2013spins}; thus they appear to disperse vertically when studied on an energy scale of $\sim$ 45 meV, as is shown here.

Both Fig. \ref{example} a) and b) also show phonon scattering, which is strongest at relatively large $Q$.  Acoustic branches are clear in Fig. \ref{example} a) as they disperse linearly from zone centres at 1,1,$L$ and 2,2,$L$ and equivalent zone centres.  The acoustic zone boundary energies are observed to be $\sim$ 14 meV.  Optic modes are observed above $\sim$ 15 meV.  Similar optic phonon scattering was better investigated in earlier studies of the LBCO system \cite{wagmanlbco,wagmanlbco2}, and the results presented here are very consistent.

The crystalline electric field excitations associated with Nd$^{3+}$ are also seen in Fig. \ref{example} a) as the almost dispersionless bands that are clear centred on $\sim$ 11 meV and 25 meV.  These are associated with neutron (and therefore dipole) induced transitions from the Kramer's doublet ground state of the $J$=9/2 multiplet appropriate to Nd$^{3+}$ to excited states at these energies.  Their presence in the Nd-LSCO system means that there are two relatively low energy windows, in which the very dispersive Cu$^{2+}$ parallel spin stripe fluctuations can be studied with little interference from the Nd$^{3+}$ CEF excitations: these are from the edge of the quasi-elastic nuclear incoherent scattering which extends to $\sim$ 4 meV and the bottom of the 1$^{st}$ excited state CEF scattering at $\sim$ 10 meV, and then from $\sim$ 14 meV to 23 meV, which lies between the scattering from the 1$^{st}$ and 2$^{nd}$ excited states of the Nd$^{3+}$ CEF scheme.

In Fig. \ref{HHL}, we explicitly examine the $L$-dependence of the inelastic scattering between 5 meV and 10 meV.  Fig. \ref{HHL} a) shows a colour contour map with integrations in energy for 5 meV $\le$ E $\le$ 10 meV, and in reciprocal space within the basal plane but normal to the $HH$ direction, -0.1 $\le$ $H\overline{H}$ $\le$ 0.1, for the T = 60 K, $x$ = 0.12 data set. The dynamic parallel spin stripes are identified as clear parallel ``ladders" of scattering along L and centred about either $HH$ = -0.5 or 0.5.  These appear extended in $L$ due to the two dimensional nature of the dynamic parallel spin stripes.

Fig. \ref{HHL} b) shows $L$-cuts through the Fig. \ref{HHL} a) colour contour map, integrating across the $HH$ direction in reciprocal space from -0.6 to -0.4 and from 0.4 to 0.6, so as to capture all of the dynamic parallel spin stripe scattering. These cuts show higher intensity at  T = 60 K as compared with  T = 5 K, which is expected due to the effect of the Bose factor on this inelastic scattering. It also shows weak structure in $L$, near $L$= 2, 4 and 6, which we attribute to weak acoustic phonons, emanating from (1, 0, $L$=2,4,6) zone centres, using orthorhombic notation (these are (1/2, 1/2, $L$) using tetragonal notation). We thus explicitly see that the dynamic parallel spin stripes have pronounced two dimensional character - no three dimensional correlations are obvious beyond those that can be attributed to weak phonon scattering.

The quality of this data also allows the clear observation of interactions between these various elementary excitations, and particularly between the 2$^{nd}$ excited state Nd$^{3+}$ CEF level and an optic phonon.  This can be seen in Fig. \ref{example} a) at [-1,-1,0], where an optic phonon at $\sim$ 17 meV disrupts the flat dispersion of the CEF excitation.  We tentatively identify this as a vibronic bound state, resulting from the entanglement of an optic phonon with the 2$^{nd}$ excited state CEF excitation of Nd$^{3+}$.  Such hybridized bound states are known to exist in other rare earth oxides, such as the spin ices Ho$_2$Ti$_2$O$_7$ and related pyrochlores\cite{jonathan}.  We also note that earlier TOF neutron scattering studies have also observed evidence for strong coupling between relatively low-lying optic phonons in the 17 meV - 19 meV range and again near 30 meV, and dispersive parallel spin stripe fluctuations in underdoped LBCO\cite{wagmanlbco,wagmanlbco2}.

TOF inelastic scattering measurements, similar to those presented in Fig. \ref{example} for Nd-LSCO $x$=0.12, were performed on single crystals of Nd-LSCO with $x$=0.19, 0.24 and 0.26, using SEQUOIA and E$_i$=60 meV.  These results, at T=5 K, are shown in panels i) - iv) and a) - p) of Fig. \ref{colorplot}.  The data in Fig. \ref{colorplot} i) - iv) has been symmetrized between [$\bar{H}$, $\bar{H}$, 0] and [$H,H,0$], while the data in Fig. \ref{colorplot} a) to p) have been symmetrized within a single quadrant of [$H,K,0$] reciprocal space.  This symmetrization overlays signal in different Brillouin zones, while averaging over any variations in background.  The intensity scale for different samples has been normalized by considering the intensity of three scattering features unrelated to Cu$^{2+}$ magnetism: the intensity of acoustic phonons in the vicinity of [1,1,0], the intensity of Nd$^{3+}$ CEF scattering near 10 meV as well as that near 25 meV.  All of these are expected to be $\sim$ independent of $x$ for Nd-LSCO with 0.12 $\le x \le$0.26.  The normalization factors used to set the relative intensities between the samples differed by $\sim$ 10$\%$, which was consistent with differences in sample masses employed.

Qualitative examination of the dynamic incommensurate spin fluctuations in the Nd-LSCO system, shows that both the ``ladder"-like spin excitations, dispersing vertically from around ($\pi$, $\pi$) wavevevectors (that is, around (0.5, 0,5, 0)) in Fig. \ref{colorplot} i) - iv), and the quartet of inelastic peaks around (0.5, 0,5, 0) in Fig. \ref{colorplot} a) - p) are present for all values of 0.12 $\le x \le$0.26.  They are somewhat sharper and easier to visually discriminate at relatively low doping, $x$=0.12, but they are present at all doping levels studied. 


\begin{figure*}[tbp]
\linespread{1}
\par
\includegraphics[width=7in,height=5in]{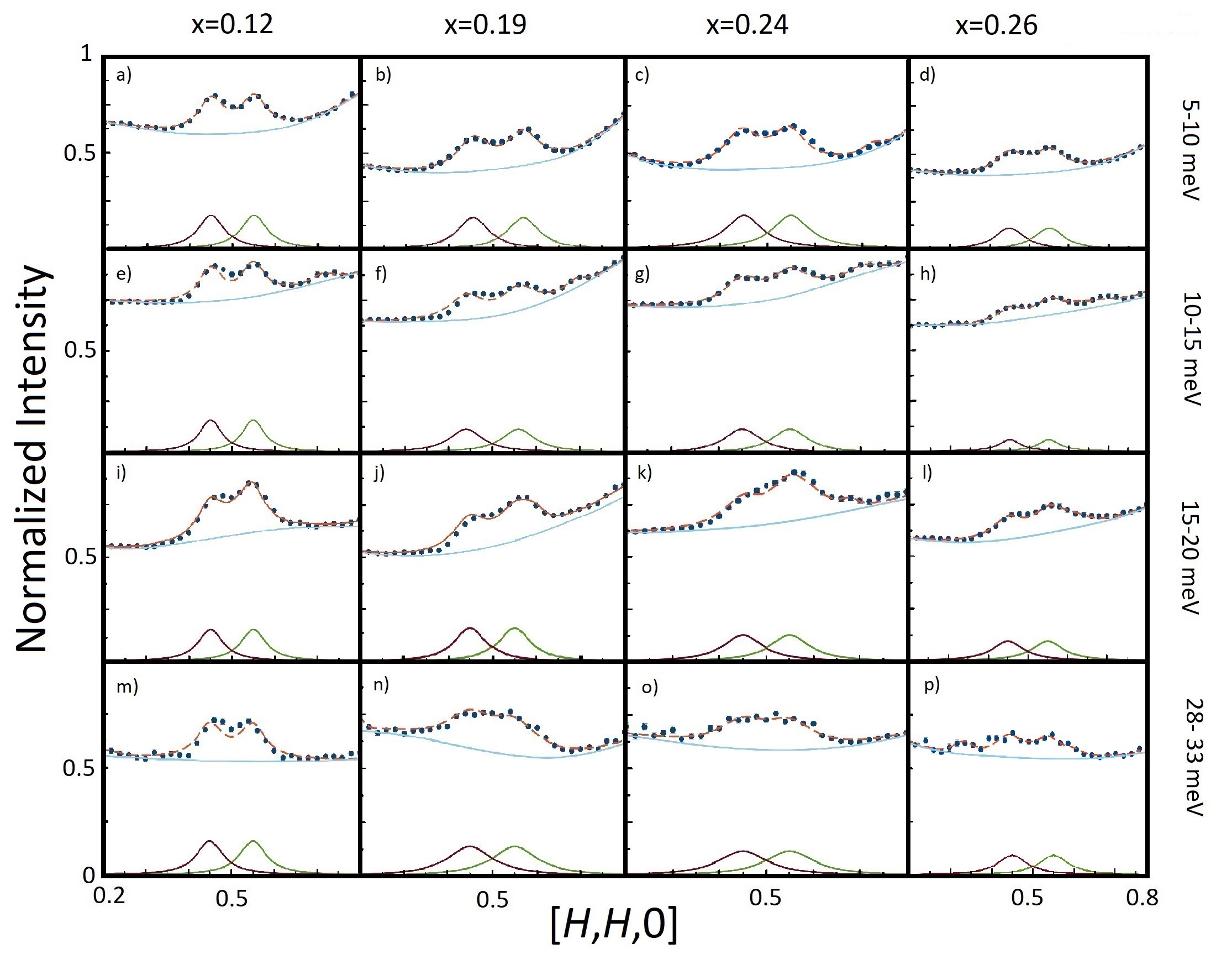}
\par
\caption{Diagonal line cuts of the inelastic neutron scattering shown in Fig. \ref{colorplot}, along the $HH$ direction in reciprocal space from 5 to 10 meV (a to d), 10 to 15 meV (e to h), 15 to 20 meV (i to l) and 28 to 33 meV (m to p). Such cuts produce two IC peaks, from the 4 IC peaks observed in Fig. \ref{colorplot} a to p. As can be seen, the signal, albeit weak, is clearly observable for all four samples $x$ = 0.12, 0.19, 0.24 and 0.26.}
\label{linecut} 
\end{figure*}

We can, of course, make a quantitative statement regarding the relative spectral weight of these parallel stripe fluctuations.  In Fig. \ref{linecut}, we present cuts along the [$H,H,0$] direction of reciprocal space through the [$H,K,0$] maps of inelastic scattering.  These cuts integrate over -4 $\le L \le$4 as well as over -0.1 to 0.1 in the [$H$,$\bar{H}$,0] direction, that is normal to the [$H,H,0$] direction in reciprocal space and within the basal plane.  Projected in this manner, the quartet of incommensurate inelastic magnetic peaks then presents as a pair of peaks symmetrically split off from the [0.5, 0.5, 0] or ($\pi$, $\pi$) wavevector.  

Figure \ref{linecut} shows such cuts over the energy range 5 meV to 10 meV, in Fig. \ref{linecut} a) - d), over the range 10 meV to 15 meV in Fig. \ref{linecut} e) - h), over the 15 meV to 20 meV in Fig. \ref{linecut} i) - l), and over the range 28 meV to 33 meV in Fig. \ref{linecut} m) - p).  Each of the four panels within these groupings corresponds to the same cut for each of the Nd-LSCO $x$=0.12, 0.19, 0.24, and 0.26 single crystal samples, as indicated in Fig. \ref{linecut}.  The $x$-dependence of the normalized relative intensity of this dynamic spectral weight is then obtained to fitting the scattering profiles in Fig. \ref{linecut} to the sum of two Lorentzians symmetrically split off from the [0.5, 0.5, 0] wavevector, with equal intensity and width and a smoothly varying background as indicated by the thin blue lines in Fig. \ref{linecut}.  In several cases an additional relatively weak Lorentzian feature was added to the background to account for structure not associated with the parallel spin stripe wavevectors.  This occurs primarily for the energy interval 10-15 meV, where the background is relatively high compared with either 5 - 10 meV or 15 - 20 meV, due to overlap with the $\sim$ 10 meV CEF excitation.  However a small additional contribution to the scattering is required in the 15 - 20 meV range, due to optic phonons.  Interference with the $\sim$ 25 meV CEF excitation makes reliable fits at higher energies in this data set difficult.  The fits to the pair of Lorentzians isolated by this procedure and indicative of the inelastic spectral weight due to Cu$^{2+}$ magnetism are shown in each panel directly below the data plus overall fits. 

\begin{figure}[tbp]
\linespread{1}
\par
\includegraphics[width=3.5in]{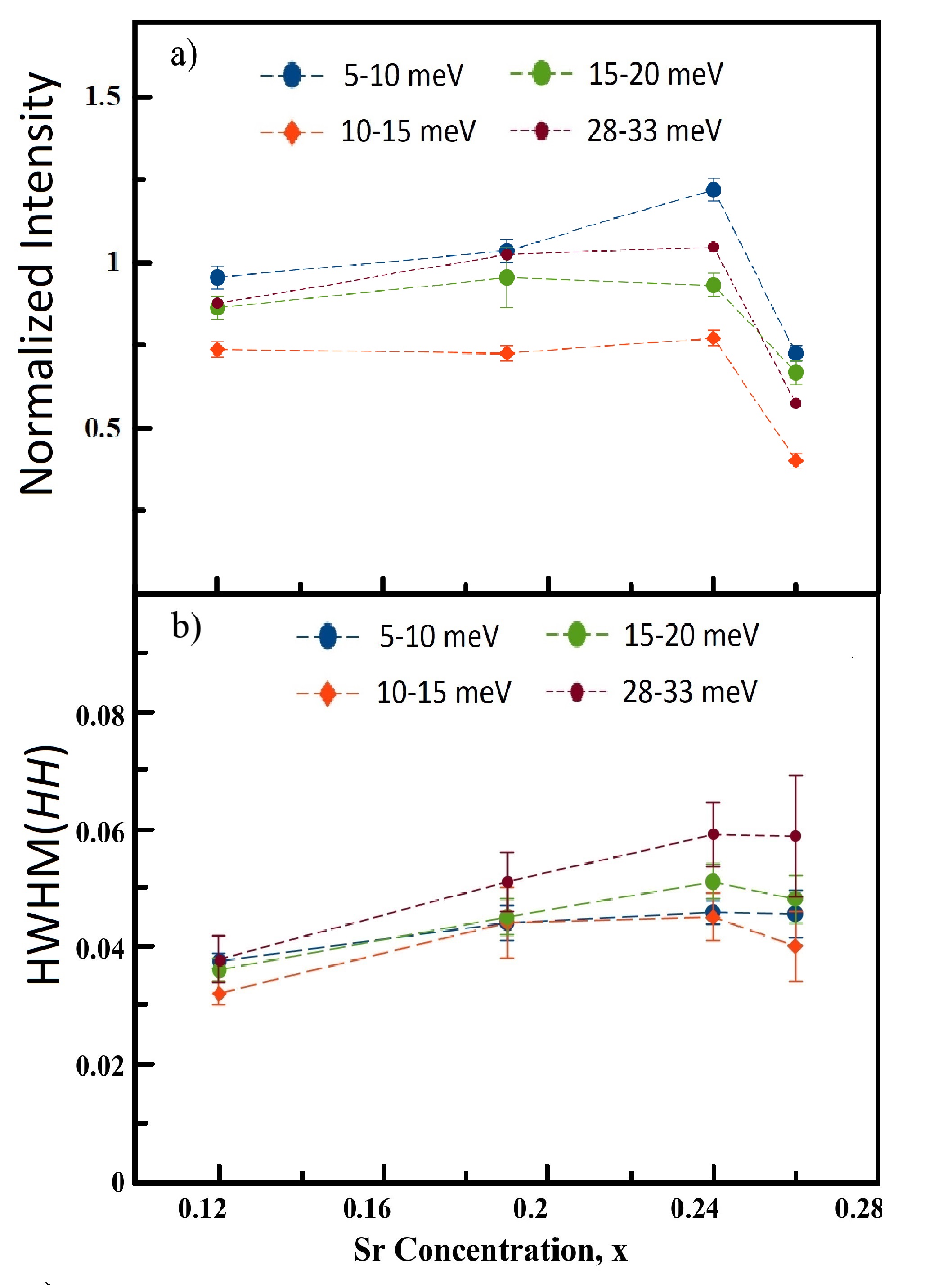}
\par
\caption{a) Neutron scattering intensities of the incommensurate peaks near ($\pi$, $\pi$) wavevectors, presented in Fig.\ref{linecut}, as a function of Sr concentration, $x$, and integrated in four energy bands: 5 to 10 meV, 10 to 15 meV, 15 to 20 meV and 28 to 33 meV. The normalized intensity evolves little across the phase diagram up to $x$=0.24 then shows a $\sim$ 35$\%$ decrease at $x$=0.26. b) The half width at half maximum (HWHM) from the fit IC peaks widths shown in Fig. \ref{linecut}.  Again the HWHM are shown for the same four energy bands as in a) and as a function of Sr concentration, $x$.}
\label{slice} 
\end{figure}

Figure  \ref{slice} a) shows the integrated dynamic spectral weight associated with dynamic parallel spin stripes in the four different energy regimes, 5 - 10 meV, 10 - 15 meV, 15 - 20 meV, and 28 - 33 meV, as a function of Sr concentration, $x$, within La$_{1.6-x}$Nd$_{0.4}$Sr$_x$CuO$_4$.  We observe minimal variation of the spectral weight of these fluctuations for all of $x$=0.12, 0.19 and 0.24.  An $\sim$ 33 $\%$ decrease in spectral weight across all energies studied is observed for Nd-LSCO with $x$=0.26, which is close to the end of the superconducting dome, and the only single crystal studied which displays the high temperature tetragonal phase as all temperatures above $T$=2K.  Figure \ref{slice} b) shows the $x$ variation in the fitted half width at half maximum (HWHM) of the inelastic Lorentzian peaks, without including the effects of instrumental resolution.  This shows a progressive broadening of the inelastic spectral weight in reciprocal space at all energies studied.  Given that the integrated dynamic spectral weight is slowly varying with a decrease at the highest hole-doping, $x$, the dynamic parallel spin stripes are quailitatively harder to see with increasing $x$, as a survey of Fig. \ref{colorplot} demonstrates.  Accounting for the momentum resolution semi-quantitatively, our current measurements are consistent with an $\sim$ 50 $\%$ increase in the width of this incommensurate inelastic scattering, consistent with the decrease in correlation length previously reported using elastic triple axis neutron scattering on the same crystals\cite{Kyle}.

\begin{figure}[tbp]
\hspace*{-0.3in}
\includegraphics[width=3.5in]{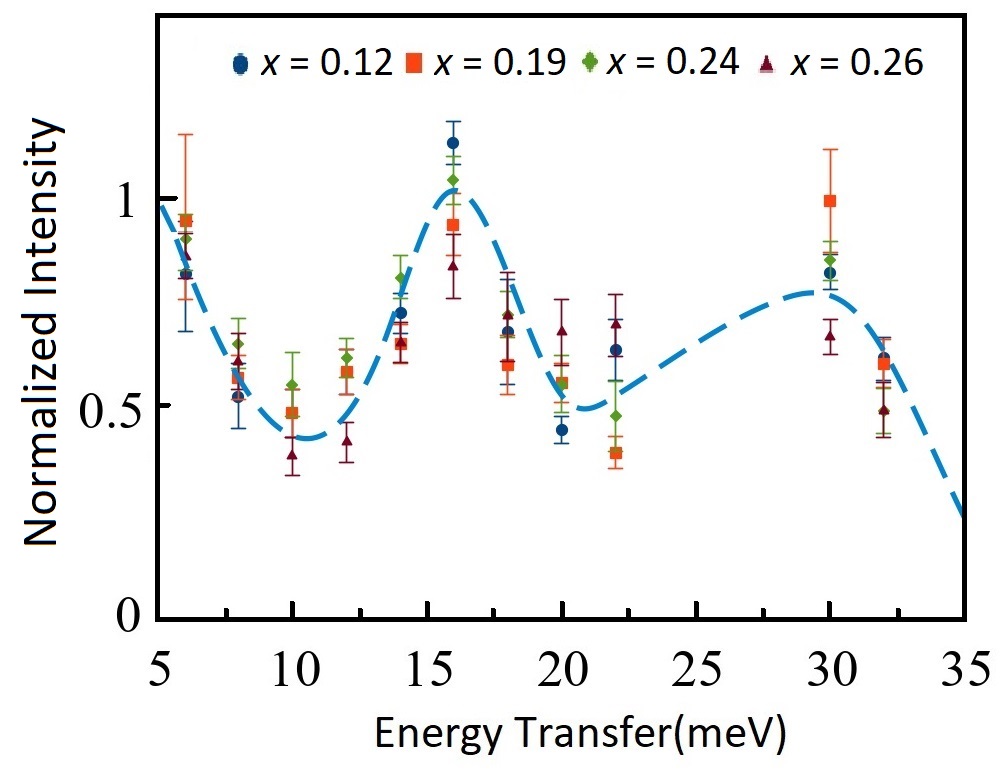}
\caption{Summary of the integrated neutron scattering intensity vs energy for all four single crystal samples, and across energy from 6meV to 33 meV. With this relatively small energy binning (2 meV) the spectral weight of the dynamic parallel spin stripes is $\sim$ $x$-independent.  The energy dependence of this spectral weight is structured with a significant peak near $\sim$ 17 meV, consistent with earlier measurements on LBCO and LSCO.} 
\label{alldata}
\end{figure}

The energy dependence of the parallel spin stripe fluctuations was investigated for the four Nd-LSCO single crystals, $x$=0.12, 0.19, 0.24 and 0.26, using the same analysis discussed above, but employing a smaller, 2 meV, energy window so as to obtain finer detail in the energy dependence of $\int S({\bf Q}, \hbar\omega) d{\bf Q}$, where the {\bf Q} integral is over the incommensurate ordering wavevectors.  Note that $\int S({\bf Q}, \hbar\omega) d{\bf Q}$=$\int \chi^{\prime\prime}({\bf Q}, \hbar\omega) d{\bf Q}$ for the range of energies ($\hbar\omega \ge$ 6 meV) and temperatures (5 K) considered here.  This is shown in Fig. \ref{alldata}, in 2 meV steps from 6 meV to 22 meV and 30 meV and 32 meV.  We observe a consistent form for this dynamic spectral weight at all $x$, with a maximum in the spectral weight near 17 meV, a fall off beyond 30 meV, and a low-energy upturn near the lowest energies measured (5 meV).  This form is largely consistent with that observed in both underdoped LBCO\cite{wagmanlbco,wagmanlbco2} and LSCO at a variety of dopings\cite{Lipscombe,vignolle2007two,LiLSCO}, where peaks in $\chi^{\prime\prime}(\hbar\omega)$ at incommensurate ordering wavevectors are observed near 15 - 17 meV and again just above 30 meV.  As discussed for LBCO\cite{wagmanlbco,wagmanlbco2}, these energies correspond to crossings of the vertically-dispersing dynamic parallel spin stripe scattering, and both zone boundary acoustic phonons, as well as optic phonons near 18 and 30 meV.  This earlier work proposed this structured energy dependence arose from hybridization of these phonons with the parallel spin stripe fluctuations.  Independent of its origin, this very distinctive energy dependence appears to be a universal feature of the 214 hole-doped cuprates, from underdoped to optimally and overdoped regimes, and the end of superconductivity. 

\section{Discussion}

Our present results show clearly that the spectral weight for dynamic parallel spin stripes around ($\pi$, $\pi$) wavevectors below $\sim$ 33 meV in Nd-LSCO change relatively little with hole doping from the 1/8 anomaly at $x$=0.12, to optimal doping at $x$=0.19, through the pseudogap quantum critical point at $x$=0.24 and $x$=0.26, the latter of which is close to the end of the superconducting dome.  This raises at least two important points for consideration: How do these results compare to the results of earlier studies on other 214 cuprate superconductors? How does these results fit into the context of theory and what is known about the evolution of the Fermi surface of these systems at relatively high doping?

Earlier inelastic scattering results from spin fluctuations in 214 cuprate superconductors present an evolving and somewhat incomplete picture\cite{wakimoto2004,wakimoto2007}. In part, this is because both inelastic neutron scattering and resonant inelastic x-ray scattering (RIXS) capabilities have themselves evolved considerably over the last decade.  In part this is also due to the fact that neutron scattering tends to concentrate on wavevectors near ($\pi$, $\pi$), where the spectral weight is strongest, while RIXS is limited to relatively small wavevectors and cannot reach in-plane wavevectors as large as ($\pi$, $\pi$). 

The earliest work examining the hole-doping, $x$, dependence of the magnetic spectral weight at low energies ($\le$ 10 meV) and near ($\pi$, $\pi$) in LSCO showed the maximum $\chi^{\prime\prime}(\omega)$ to scale with superconducting $T_C$ for 0.25$\le$ x$\le 0.3$\cite{wakimoto2004}.  This result implied that the magnetic spectral weight falls off relatively strongly with doping, at least at high doping. Later neutron work extended this to higher energies with a similar conclusion, with a large difference in dynamic spectral weight up to $\sim$ 100 meV reported between LBCO at $x$=0.125 and LSCO at x=0.25 and 0.3\cite{wakimoto2007}.  However RIXS measurements on epitaxial thin films, first in the ($H,0$) direction of in-plane reciprocal space\cite{dean2012,dean2013persistence}, and later in the ($H,H$) direction\cite{meyers}, both reported high energy``paramagnon" scattering with relatively little x-dependence to its integrated spectral weight over a broad range of hole-doping in LSCO, 0$\le$ x$\le$ 0.4 for the ($H,0$) direction and 0$\le x \le$ 0.26 for the ($H,H$) direction. The long wavelength of the incident x-rays employed at the Cu L$_3$ edge in these RIXS measurements restricted the largest wavevector accessed in the ($H,H$) direction to be just beyond (0.3, 0.3) and well removed from the ($\pi$, $\pi$) position, (0.5, 0.5). Work by Wakimoto $et \: al$ in 2015 \cite{WakimotoRIXS} and Monney $et \: al$ in 2016 \cite{monney2016} showed that while high energy neutron scattering and RIXS scattering covered complementary regions of the Brillouin zone, the ``paramagnon" {\it dispersion} measured with both techniques was largely consistent, and closely resembled that of the 214 parent material La$_2$CuO$_4$ in each of the ($H$,$H$) and ($H$,0) in-plane directions in reciprocal space. Later RIXS measurements \cite{Robarts2019} on LSCO single crystals at $x$=0, 0.12 and 0.16, show an initial increase in "paramagnon" spectral weight at both (0.25, 0) and (0.25, 0.25) and up to $\sim$ 400 meV from $x$=0 to $x$=0.12, with little change between $x$=0.12 and 0.16. 

Our results on the persistence of the spectral weight for dynamic parallel spin stripes, that is magnetic fluctuations near ($\pi$, $\pi$) wavevectors, and below $\sim$ 33 meV in Nd-LSCO appear inconsistent with earlier neutron results on LSCO, as discussed above.  It is possible that the incommensurate magnetism in LSCO and LBCO is simply different than that in Nd-LSCO, although this seems unlikely given the strong similarities between all the 214 cuprate families.  The estimate of the normalized intensities of these magnetic excitations becomes progressively more difficult with increasing $x$, as the widths of these excitations becomes broader, and they are therefore harder to distinguish from both background and other excitations, such as phonons.  Better neutron instrumentation, as is being continuously developed, will no doubt help resolve these issues going forward.  However, our measurements showing the persistence of the dynamic parallel spin stripes over this broad range of doping is consistent with the results of all RIXS measurements reported to date on LSCO, at in plane wavevectors both along ($HH$) and ($H0$).  

Several recent works have focused on theoretical models relevant to the cuprates near optimal and high doping [\cite{hubbard,li2021superconductor,schackleton2021}. Two of these allow for concrete comparisons with the present neutron scattering work on Nd-LSCO.  

A recent theoretical study has examined the Hubbard model on a two dimensional square lattice and both near-neighbour and next-near neighbour hopping, as well as an on-site Coulomb interaction\cite{hubbard}.  This model can account for both a Lifshitz transition at high doping, and a transition to an electron-like Fermi surface at $p^*$.  As a function  of hole doping, this theory shows an evolution in $\int \chi^{\prime\prime} d\omega$ from concentrated around ($\pi$, $\pi$) at $x$=0.15 to more evenly distributed throughout the Brillouin zone at $x$=0.25.  At low energies $\chi^{\prime\prime}({\bf Q}, \hbar\omega)$ shows strong x-dependence at ($\pi$, $\pi$), but a much weaker $x$-dependence at ($\pi$, 0) and (0, 0).

The work by Shackleton $et \: al$ \cite{schackleton2021} is particularly interesting.  It presents finite-cluster exact diagonalization of a t-J model for $S$=1/2 electrons with random all-to-all hoping, and hence can be thought of as a model for a quantum spin glass.  At zero temperature, it shows a spin glass phase with static order out to $p_C$ $\sim$ 0.33, similar to the observed behaviour of quasi-Bragg peaks in Nd-LSCO out to at least $x$=0.26, but with dynamic spectral weight which is $\sim$ $x$-independent over a large range of dopings, similar to that reported here. 

One conclusion which seems clear is that the persistence of the dynamic parallel spin stripes which we observe in Nd-LSCO is inconsistent with models of the inelastic incommensurate magnetic scattering arising from nesting of portions of the Fermi surface.  Such an origin for the dynamic parallel spin stripes would imply substantial changes in this scattering across $p^*$=$x^*$=0.23 in Nd-LSCO, as Hall measurements in Nd-LSCO show a pronounced change in the $x$-dependence of the Hall number, from $p$ to 1+$p$, implying a significant change in Fermi surface properties at $p^*$. ARPES measurements on the Nd-LSCO system by C.E.Matt \textit{et al}\cite{Mattarpes2015} have also shown the closing of an antinodal pseudogap at $\sim$ $p^*$. We therefore believe that a local spin nature to the incommensurate spin fluctuations is required to account for such persistence, making them vestiges of the parent insulator.  This persistence of the dynamic parallel spin stripes we observe is captured by the quantum spin glass, random t-J model\cite{schackleton2021}, but does not appear to be accounted for in detail within the aforementioned Hubbard model theory\cite{hubbard}, which predicts relatively strong reduction in $\chi^{\prime\prime}({\bf Q}, \hbar\omega)$ at low $\hbar\omega$, and at ($\pi$, $\pi$) as a function of $x$. However, the overall $x$ dependence of the calculated $\int \int \chi^{\prime\prime} d\omega d{\bf Q}$, integrated in {\bf Q} over the Brillouin zone, is likely much more modest, and could be more consistent with the conclusions for the persistence of dynamic parallel spin stripes we present here.

\section{Conclusions}

We have carried out TOF inelastic neutron scattering measurements on single crystals of Nd-LSCO with hole doping, $x$, ranging from the 1/8 anomaly at $x$=0.12 to optimal doping at $x$=0.19, to just beyond the pseudogap quantum critical point at $x$=0.24, to the end of the superconducting dome, at $x$=0.26.  For all four dopings, we observe structured spectral weight at incommensurate wavevectors around ($\pi$, $\pi$), which peaks at $\sim$ 17 meV, consistent with earlier measurements on both LBCO \cite{wagmanlbco,wagmanlbco2} and LSCO \cite{Lipscombe,vignolle2007two,LiLSCO}.

Our main conclusion is that this incommensurate magnetic spectral weight near ($\pi$, $\pi$), which we refer to as dynamic parallel spin stripes, is $\sim$ independent of hole-doping, $x$, for 0.12 $\le x \le$0.24, and then displays an $\sim$ 33 $\%$ drop at $x$=0.26, as superconductivity recedes.  This observed persistence of the low energy ($<$ 33 meV) spin fluctuations in Nd-LSCO with doping is consistent with the persistence of relatively high energy ``paramagnon" scattering reported in LSCO using RIXS techniques \cite{dean2013persistence}.

Recent elastic neutron scattering studies on the same four single crystal Nd-LSCO samples showed ``static" parallel spin stripe order in the form of quasi-Bragg peaks to extend out to the end of the superconducting dome, albeit with diminishing ordered moment and in plane correlation lengths\cite{Kyle}.  The present inelastic measurements reach a similar conclusion regarding dynamic parallel spin stripe fluctuations: that they persist across the superconducting dome, with largely undiminished spectral weight, until the end of the superconducting dome, and continue to play a role in the physics of superconducting hole-doped cuprates at high doping.  Both of these features are reproduced within the recent quantum spin glass, random $t$-$J$ model\cite{schackleton2021}.  We hope that this study helps to guide future studies, both theory and experiment, such that the intertwined nature of the superconducting and magnetic states of matter as expressed in the cuprates can be fully understood.

\section{Acknowledgments}

We thank Amireza Ataei, Nigel Hussey, Takashi Imai, Hae-Young Kee, Steven Kivelson, Graeme Luke, Subir Sachdev, Louis Taillefer, and John Tranquada for useful and stimulating discussions. This work was supported by the Natural Sciences and Engineering Research Council of Canada. This research used resources at the Spallation Neutron Source, DOE Office of Science User Facilities operated by the Oak Ridge National Laboratory (ORNL).

\section{Appendix A: Alternative, Gaussian lineshape description of the constant-energy, [$H,H$,0] scans of the dynamic parallel spin stripes}
\begin{figure}[tbp]
\hspace*{-0.2in}
\includegraphics[width=3.8in]{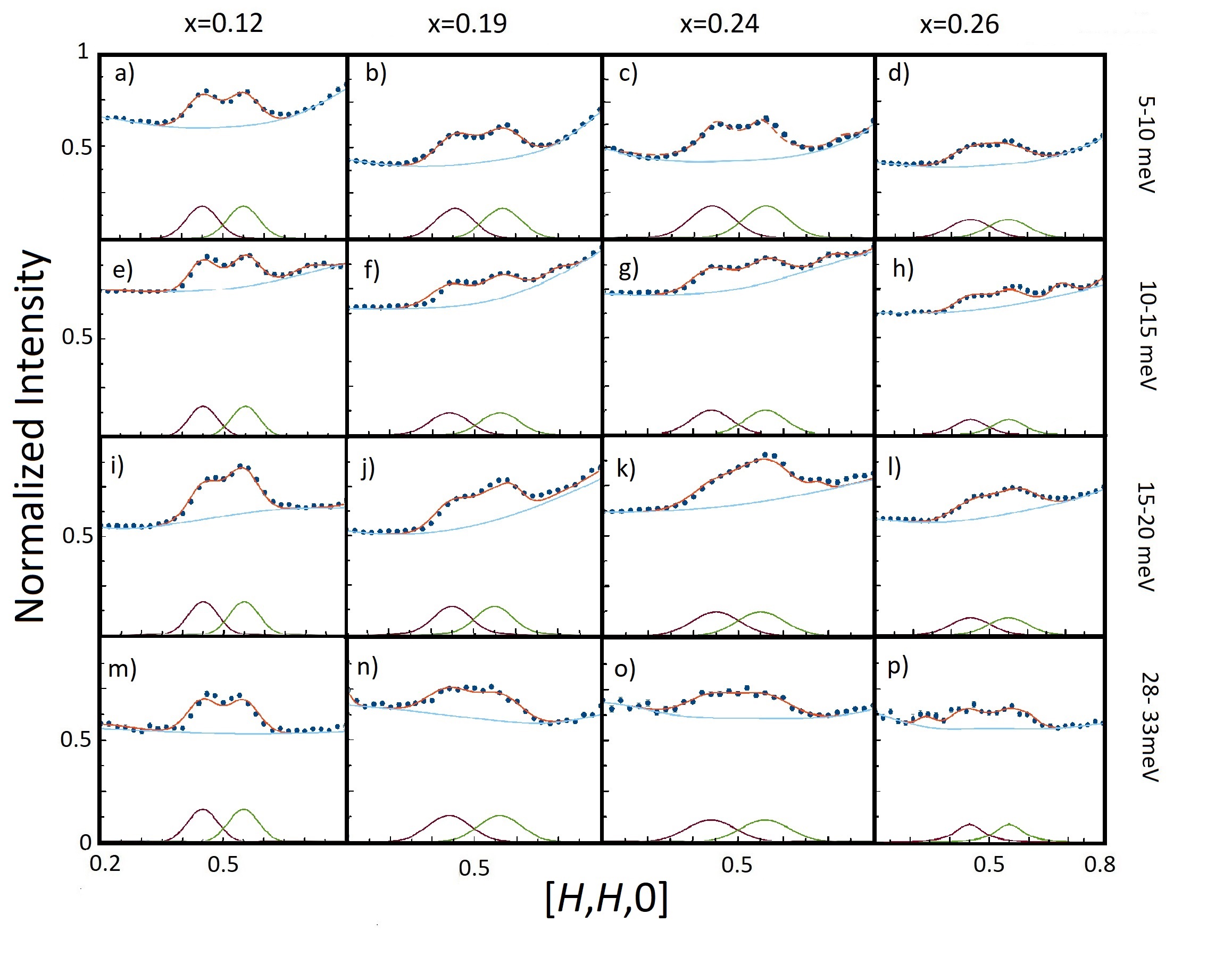}
\caption{Diagonal line cuts of the inelastic neutron scattering shown in Fig. \ref{colorplot}, along the $HH$ direction in reciprocal space from 5 to 10 meV (a to d), 10 to 15 meV (e to h), 15 to 20 meV (i to l) and 28 to 33 meV (m to p). The IC peaks are now fitted with Gaussian lineshapes, as opposed to the Lorentzian lineshapes employed in Fig.\ref{linecut}.}
\label{linecutgau} 
\end{figure}

\begin{figure}[tbp]
\vspace*{0.1in}
\linespread{1}
\par
\includegraphics[width=3in]{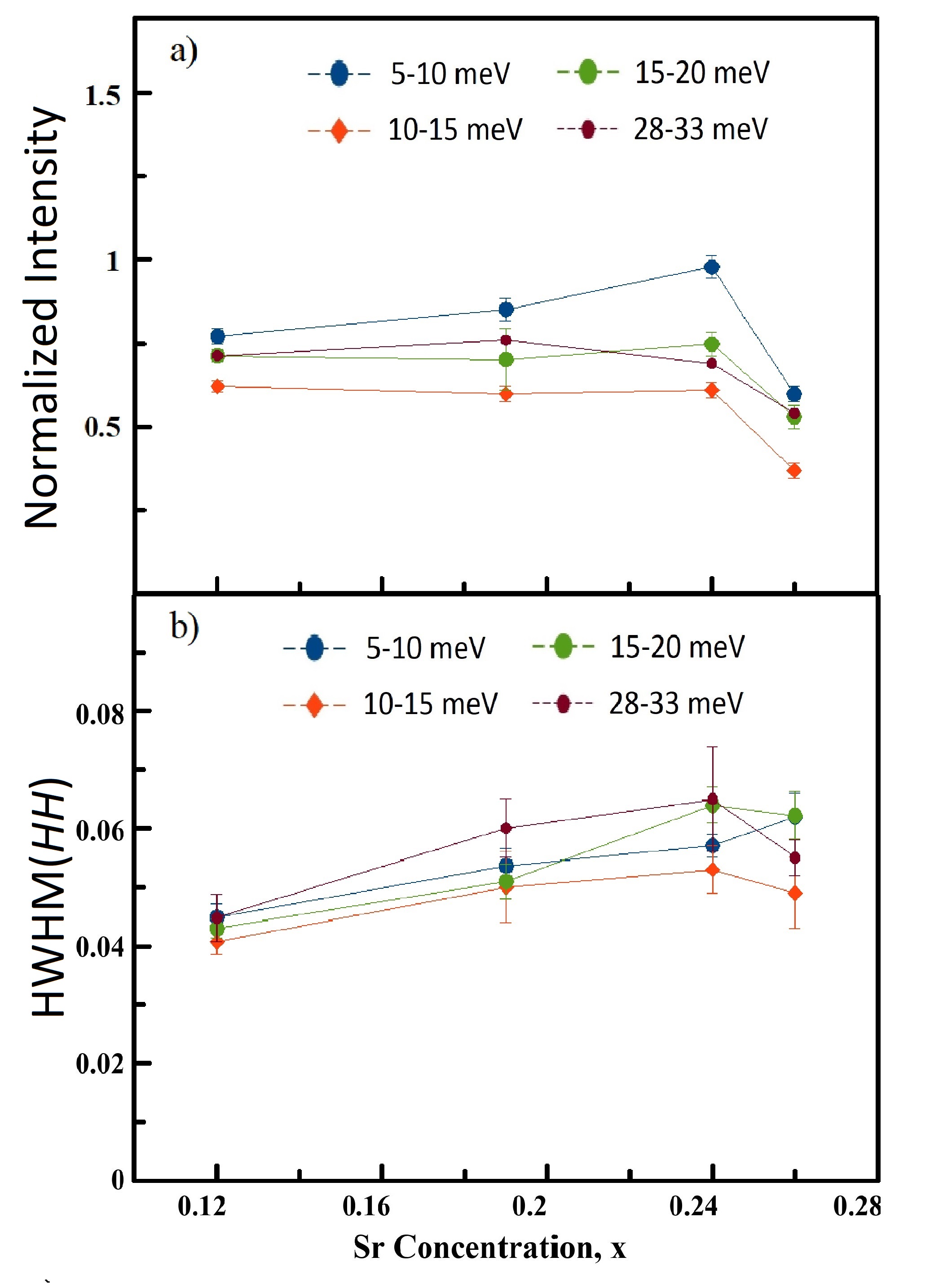}
\par
\caption{a) Normalized neutron scattering intensities of the incommensurate peaks near ($\pi$, $\pi$) wavevectors, presented in Fig.\ref{linecutgau}, as a function of Sr concentration, $x$, and integrated in four energy bands b) The half width at half maximum (HWHM) from the fit IC peaks widths shown in Fig. \ref{linecutgau}. All data are obtained using Gaussian lineshapes, as opposed to the Lorentzian lineshapes employed in Fig. \ref{slice}.}
\label{slicegau} 
\end{figure}

In the section III and Fig. \ref{linecut}, we employed a fitting algorithm of constant-energy cuts of the inelastic neutron scattering data which used Lorentzian functions to describe the inelastic, IC magnetic peaks near the ($\pi$, $\pi$) positions in reciprocal space.  The resulting normalized intensities and HWHM of the fitted peaks were plotted in Fig. \ref{slice} which showed little variation of the IC peak intensities across x=0.12, 0.19 and 0.24, with an $\sim$ 33 $\%$ drop in spectral weight for x=0.26. In the interest of investigating the sensitivity of these results to the inelastic IC lineshape employed, we repeated the analysis leading to the fittings shown in Fig. \ref{linecut}, but using Gaussian lineshapes, rather than Lorentzian lineshapes to describe the IC inelastic peaks.  The primary difference between the two is that the Gaussian lineshapes have much restricted tails, as compared with Lorentzian.  The results of this alternative fitting protocol is shown in Fig. \ref{linecutgau}, which also results in a good quality fit to the data.  The resulting x-dependence of the normalized integrated intensities of the fitted IC inelastic scattering, as well as the corresponding HWHM are shown in Fig. \ref{slicegau}.  As can be seen, the trend of the x-dependence of both the spectral weight and the HWHM are very similar to the results shown in Fig. \ref{slice}, and we conclude that our results regarding the x-dependence of the inelastic IC spectral weight, that is the persistence of the strength of the dynamic parallel spin stripes across 0.12 $\le$ $x$ $\le$ 0.26 are robust to the details of the choice of function used to describe them.

\FloatBarrier


%

\end{document}